\documentclass[%
 reprint,
 amsmath,amssymb,
 aps,
 pra,
floatfix,
]{revtex4-2}

\usepackage{graphicx}
\usepackage{dcolumn}
\usepackage{bm}
\usepackage{orcidlink}
\usepackage{array}
\usepackage{hyperref}

\begin{document}

\title{Stochastic perturbation theory: a prequel to Reptation Quantum Monte Carlo}
\author{Stefano Baroni\,\orcidlink{0000-0002-3508-6663}}
\affiliation{SISSA --  Scuola Internazionale Superiore di Studi Avanzati, I-34136 Trieste (European Union)}
\affiliation{CNR, Istituto dell'Officina dei Materiali, SISSA unit, I-34136 Trieste (European Union)}

\begin{abstract}
  I present a different approach to Rayleigh-Schr\"odinger perturbation theory, based on Laplace transforms and polynomial theory, yielding an iterative expression for the perturbative expansion of the energy of the non-degenerate ground state of a quantum system, which easily lends itself to symbolic computation. A stochastic interpretation of the various perturbative corrections naturally leads to a re-summation scheme that is equivalent to Reptation Quantum Monte Carlo and that actually provided the original motivation to its development in the late nineties.
\end{abstract}

\maketitle

\section{Introduction} \label{Sec:Introduction}
Perturbation theory (PT) \cite{Picasso2014} is as old as modern quantum mechanics (QM) itself \cite{Schrodinger:1926c}, and is in fact one of the pillars of any elementary or advanced course in QM. PT is instrumental to most applications of QM, other than a few exactly solvable models, and has provided the ground for advanced methods, such as quantum field theory in particle and condensed-matter physics, or quantum chemistry. In spite of its ubiquity, the use of PT is restricted to low orders, for its complexity increases very steeply with the order of the theory. Non-perturbative methods, such as those based on stochastic approaches, have therefore gained popularity due to their broader applicability.

The purpose of this paper is twofold. On the one hand, it presents a novel approach to PT, based on Laplace transforms and polynomial theory, that allows perturbative corrections to the ground-state (GS) energy of a quantum system to be derived to any order, without ever computing any corrections to the wavefunction. While this approach hardly broadens the scope of PT, it does provide a systematic and mathematically elegant approach to it, which easily lends itself to automatic algebraic manipulation. On the other hand, a well established mapping between the imaginary-time evolution of a quantum system and the diffusive process of an auxiliary classical system \cite{Parisi:1981} allows one to interpret the perturbative corrections as cumulants of a suitably defined random walk and suggests a re-summation scheme, which is equivalent to Reptation Quantum Monte Carlo (RQMC) \cite{Baroni:1999a,Baroni:1999b} and that actually provided the original motivation to its development in the late nineties.

This paper is organized as follows: Sec. \ref{sec:RalSchr} presents a new approach to Rayleigh-Schr\"odinger PT, not requiring the calculation of any corrections to the wavefunction; Sec. \ref{sec:ClaQua} introduces the quantum-classical mapping that is propedeutic to stochastic perturbation theory and RQMC; Secs. \ref{sec:StochPert} and \ref{sec:RQMC} present a stochastic interpretation of PT theory and RQMC as an effective technique to resum all the perturbative corrections up to infinite order; finally, Sec. \ref{sec:Conclusions} contains my conclusions.

\medskip\section{A different path to Rayleigh- Schr\"odinger Perturbation theory} \label{sec:RalSchr}

We want to compute the GS energy, $E_0$, of a quantum system whose the Hamiltonian, $\widehat H$, can be split into an unperturbed term, $\widehat {\mathcal H}$, whose spectrum is known,
\begin{equation} \label{eq:eigenvalue}
  \widehat{\mathcal{H}}\Phi_{n}=\mathcal{E}_{n}\Phi_{n},
\end{equation}
and a perturbation, $\widehat{\mathcal W}$:
\begin{equation}
  \widehat{H} = \mathcal{\widehat{H}}+\widehat{\mathcal{W}}. \label{eq:H-split}
\end{equation}
The purpose of perturbation theory is to express $E_0$ as a power series in the stregth of the perturbation, $\mathcal W$. In order to streamline some of the notation, I will assume that the energy zero is chosen to coincide with the unperturbed ground state: $\mathcal{E}_{0}=0$. If the latter is not orthogonal to the exact one, one has:
\begin{equation}
  \begin{gathered}
    E_{0} \sim -\frac{d }{d\tau}\cal \log\cal Z(\tau), \\
    {\cal Z}(\tau)  = \langle\Phi_{0}|\mathrm{e}^{-\widehat{H}\tau}|\Phi_{0}\rangle = \sum_{n=0}^\infty |\langle\Phi_0|\Psi_n\rangle|^2 \mathrm{e}^{- E_n\tau},
  \end{gathered} \label{eq:E-dlog}
\end{equation}
where $E_n$, and $\Psi_n$ are eigenpairs of the exact Hamiltonian, $\widehat H$,
\begin{widetext}
\begin{multline}
  \mathrm{e}^{-\widehat{H}\tau}=\mathrm{e}^{-\widehat{\mathcal{H}}\tau}\left(1-\int_{0}^{\tau}d\tau_{1}\widehat{\mathcal{W}}(\tau_{1})+\int_{0}^{\tau}d\tau_{2}\int_{0}^{\tau_{2}}d\tau_{1}
  \widehat{\mathcal{W}}(\tau_{2})\widehat{\mathcal{W}}(\tau_{1})+\right. \\ \left. \cdots(-)^{n} \int_{0}^{\tau}d\tau_{n} \int_{0}^{\tau_{n}} d\tau_{n-1} \cdots\ \int_{0}^{\tau_{2}}
  d\tau_{1} \widehat{\mathcal{W}}(\tau_{n}) \widehat{\mathcal{W}}(\tau_{n-1}) \cdots\widehat{\mathcal{W}}(\tau_{1})+\cdots\right),
\end{multline}
its the imaginary-time propagator, $\widehat{\mathcal{W}}(t) =\mathrm{e}^{\widehat{\mathcal{H}}\tau} \widehat{\mathcal{W}}\mathrm{e}^{-\widehat{\mathcal{H}}\tau}$ is the perturbation in the interaction representation,
the ``$\sim$'' symbol indicates the large (imaginary-) time limit, and natural units ($\hbar=1$) are  used throughout this paper.
%
We can thus write a perturbative expansion for $\mathcal Z(\tau)$ as:
\begin{equation} \label{eq:Zdef}
  \mathcal Z(\tau) =1-\lambda_{1}(\tau) + \lambda_{2}(\tau) + \cdots(-)^{n} \lambda_{n}(\tau)+\cdots,
\end{equation}
where $\lambda_1(\tau)=\mathcal W_{00}$ and the $n$-th order term ($n\ne 0$) reads:
\begin{align}
  \lambda_{n}(\tau) &  =\int_{0}^{\tau}d\tau_{n}\int_{0}^{\tau_{n}}d\tau_{n-1}\cdots\int_{0}^{\tau_{2}}d\tau_{1}\langle\Phi_{0}|\widehat{\mathcal{W}}(\tau_{n})
    \widehat{\mathcal{W}}(\tau_{n-1})\cdots\widehat{\mathcal{W}}(\tau_{1})|\Phi_{0}\rangle \label{eq:lambda-def} \\
    & =
      \sum_{k_{1}\cdots k_{n-1}}\mathcal{W}_{0k_{n-1}}\mathcal{W}_{k_{n-1}k_{n-2}} \cdots \mathcal{W}_{k_10}\int_{0}^{\tau}d\tau_{n}
      \int_{0}^{\tau_{n}}d\tau_{n-1}\mathrm{e}^{-\mathcal{E}_{k_{n-1}}(\tau_{n}-\tau_{n-1})}
      \cdots\int_{0}^{\tau_{2}}d\tau_{1}\mathrm{e}^{-\mathcal{E}_{k_{1}}(\tau_{2}-\tau_{1})},
    \label{eq:lambda-Sum_k}
\end{align}
\end{widetext}
and $\mathcal W_{kl}=\langle\Phi_k|\widehat{\mathcal W}|\Phi_l \rangle $. Note that the large-time behaviour of $\lambda_n(\tau)$ is polynomial, of order $n$: $ \lambda_n(\tau) \sim \mathcal O ( \tau^n )$. In order to express $\log\mathcal Z$ as a power series in the strength of the perturbation, $\mathcal W$, we define the formal moments as: $\mu_n=n! \lambda_n$. The logarithm of $\mathcal Z$ can then be expressed as a power series in $\mathcal W$ as:
\begin{equation}
    -\log \mathcal Z(\tau) =\kappa_{1}(\tau)-\frac{1}{2}\kappa_2(\tau) \cdots + \frac{(-)^{n+1}}{n!}\kappa_{n}(\tau) \cdots,
  \label{eq:cumulant-expansion}
\end{equation}
where the formal cumulants, $\kappa_n$, are defined as \cite{withers:2009}:
\begin{equation}
  \begin{aligned}
    \kappa_{1} & =\mu_{1} \\
    \kappa_{2} & =\mu_{2}-\mu_{1}^{2} \\
    \kappa_{3} & =\mu_{3}-3\mu_{2}\mu_{1}+2\mu_{1}^{3} \\
    & \cdots \\
    \kappa_{n} & =\mu_{n}-\sum_{k=1}^{n-1}{n-1 \choose k}\kappa_{n-k}\mu_{k}.
  \end{aligned}
  \label{eq:kappa-recursion}
\end{equation}
The recursive relation between moments and cumulants, Eq. \eqref{eq:kappa-recursion}, is best expressed in terms of reduced cumulants, $\gamma_n=\kappa_n/n!$ as:
\begin{equation}
    \gamma_n(\tau) = \lambda_n(\tau) -\sum_{k=1}^{n-1} \frac{n-k}{n}\gamma_{n-k}(\tau)\lambda_k(\tau). \label{eq:gamma-recursion}
\end{equation}
We thus have:
\begin{equation}
  \begin{aligned}
    E_{0} &= \varepsilon_1 + \varepsilon_2 + \cdots \varepsilon_n+ \cdots, \\
    \varepsilon_n &\sim (-)^{n+1}\dot\gamma_{n}(\tau),
  \end{aligned}
     \label{eq:kappa-perturbative}
\end{equation}
where $\varepsilon_n$ is the $n$-th order correction and the dot indicates a derivative with respect to imaginary time.

In order for the limit implicit in Eq. \eqref{eq:kappa-perturbative} to exist, it is necessary that the $\kappa$'s grow at most linearly with $\tau$ as $\tau\to\infty$. I do not know how this property can be demonstrated, other than from the tautology that the limit \emph{must} exist. In Sec. \ref{sec:StochPert}, where perturbation theory will be expressed in terms of an auxiliary stochastic process, eventually leading to RQMC, this property will be shown to derive from the additivity of the cumulants of sums of independent stochastic variables.

Using Eq. \eqref{eq:gamma-recursion}, a recursion relation can be written for the $\dot\gamma$'s in terms of the $\lambda$'s and their derivatives:
\begin{multline}
    \dot{\gamma}_{n}(\tau)=\dot{\lambda}_{n}(\tau)- \\ \sum_{k=1}^{n-1}\frac{n-k}{n}\left(\dot{\gamma}_{n-k}(\tau)\lambda_{k}(\tau)+\gamma_{n-k}(\tau)\dot{\lambda}_{k}(\tau)\right).
  \label{eq:gamma-dot-recursion}
\end{multline}
The left-hand side of Eq. \eqref{eq:gamma-dot-recursion} is $\sim \mathcal{O}(1)$, whereas the right-hand side features terms of orders up to $\sim \mathcal{O}(\tau^{n-1})$, which cancel out each other and would be wasteful to compute. In order to dash off the discussion to follow, I denote by $x^{\circ}$ the term of order zero, $\sim \mathcal{O}(1)$, in the asymptotic expansion of $x(\tau)$ as $\tau\to\infty$. Of course,
$\dot x^\circ$ indicates the zero-th order term of $\dot x(\tau)$ and not the derivative of the zero-th order term, which would otherwise vanish. Eqs. \eqref{eq:gamma-recursion} and \eqref{eq:gamma-dot-recursion} hold verbatim for the values of the constant terms in the asymptotic expansions of $\gamma_n(\tau)$ and $\dot\gamma_n(\tau)$, $\gamma_n^\circ$
and $\dot\gamma_n^\circ$---the latter coinciding with the $\tau\to\infty$ limit---in terms of the $\lambda^\circ$'s and $\dot\lambda^\circ$'s:
\begin{equation} \label{eq:gamma-naught-recursion}
  \begin{aligned}
      \gamma^\circ_n &= \lambda^\circ_n -\sum_{k=1}^{n-1} \frac{n-k}{n}\gamma^\circ_{n-k}\lambda^\circ_k,  \\
      {\dot \gamma}^\circ_n&=\dot{\lambda}_{n}^{\circ}-\sum_{k=1}^{n-1}\frac{n-k}{n}\left(\dot{\gamma}_{n-k}^{\circ}\lambda_{k}^{\circ}+\gamma_{n-k}^{\circ}\dot{\lambda}_{k}^{\circ}\right).
    \end{aligned}
\end{equation}

The asymptotic ($\tau\to\infty$) behaviour of a function of a real argument, such as $\lambda_n(\tau)$, is determined by the analytical properties of its Laplace transform,
\begin{equation}
    \bar{\lambda}_{n}(z) \doteq\int_{0}^{\infty}\lambda_{n}(\tau)\mathrm{e}^{-z\tau}d\tau, \label{eq:lambda-bar}
\end{equation}
near the origin, $z=0$. In fact, as the Laplace transform of $\tau^n$ is $n!/z^{n+1}$, $\lambda^\circ_n$ and $\dot\lambda^\circ_n$ are the coefficients of order $-1$ and $-2$, respectively, of the Laurent expansion of $\bar\lambda_n(z)$ around the origin. In order to evaluate Eq. \eqref{eq:lambda-bar},
we note that the multiple integral in Eq. \eqref{eq:lambda-Sum_k} is the convolution: $1\ast  \mathrm{e}^{-\mathcal E_{k_{n-1}}\tau} \cdots \ast \mathrm{e}^{-\mathcal E_{k_1}\tau} \ast 1$, whose Laplace transform is: $1/(\mathcal E_{k_{n-1}}+z) \cdots / (\mathcal E_{k_1}+z)/z^2$.
Therefore,
\begin{equation}
    \bar{\lambda}_{n}(z) =\frac{1}{z^{2}}G_{n}(z),
\end{equation}
where $G_1(z)=\mathcal W_{00}$ and for $n>1$
\begin{align}
    G_{n}(z) =
        \sum_{k_{1}\cdots k_{n-1}}\frac{\mathcal{W}_{0k_{n-1}}\mathcal{W}_{k_{n-1}k_{n-2}}\cdots\mathcal{W}_{k_10}}{(\mathcal{E}_{k_{n-1}}+z)
        \cdots(\mathcal{E}_{k_{1}}+z)}.
        \label{eq:Gn}
\end{align}
We conclude that $\lambda^\circ_n$ and $\dot\lambda^\circ_n$ are the coefficients of order one and zero, respectively, in the Laurent expansion of $G_n(z)$ around the origin. For future reference, it is expedient to designate the term where no ground-state contributions to the sum in Eq. \eqref{eq:Gn} occur as:
\begin{align}
  g_{n}(z)  =
      \sideset{}{'}\sum_{k_{1}k_{2}\cdots k_{n-1}}
      \frac{\mathcal{W}_{0k_{n-1}}
      \mathcal{W}_{k_{n-1}k_{n-2}}\cdots\mathcal{W}_{k_10}}{(\mathcal{E}_{k_{n-1}}+z)\cdots(\mathcal{E}_{k_{1}}+z)},
    \label{eq:gn}
\end{align}
where $\sideset{}{'}\sum$ indicates a multiple sum excluding all the terms where at least on the indices vanishes, $k_{i}=0$.

The analytical behaviour of the various terms appearing in Eq. \eqref{eq:Gn} is determined by the number of times the ground state ($k_i=0$) occurs in each one of them, each time raising the order of the pole at $z=0$ by one unit. Let us depict any such term as a sequence of $n+1$ boxes, each labeled by a summation index, $k_i$, with the two indices at the extrema being kept equal to zero, $k_0=k_n=0$:
\setbox0\hbox{$k_{n-1}$}
\newlength{\cwd}\setlength{\cwd}{\wd0}
\begin{center}
    \begin{tabular}{ |>{\centering\arraybackslash}m{\cwd}|>{\centering\arraybackslash}m{\cwd}|>{\centering\arraybackslash}m{\cwd}|>{\centering\arraybackslash}m{\cwd}|>{\centering\arraybackslash}m{\cwd}|>{\centering\arraybackslash}m{\cwd}| }
    \hline
    0 & $k_1$ & $k_2$ & $\cdots$ & $k_{n-1}$ & 0 \\
    \hline
  \end{tabular}~.
\end{center}
We can now partition Eq. \eqref{eq:Gn} into partial sums, each one characterized by the number $\ell$ of vanishing $k_i$ indices ($\ell=0,\cdots n-1$). Any term of a partial sum is the ratio between the product of $\ell+1$ $g$'s (Eq. \ref{eq:gn}), which is a regular function as $z\to 0$, and $z^\ell$. For instance, one term of the $\ell=2$ partial sum could look like:
\begin{center}
    \smallskip
     \begin{tabular}{
     |>{\centering\arraybackslash}m{\cwd}
     |>{\centering\arraybackslash}m{\cwd}
     |>{\centering\arraybackslash}m{\cwd}
     |>{\centering\arraybackslash}m{\cwd}
     |>{\centering\arraybackslash}m{\cwd}
     |>{\centering\arraybackslash}m{\cwd}
     |>{\centering\arraybackslash}m{\cwd}
     |>{\centering\arraybackslash}m{\cwd}
     | c }
    \cline{1-8}
    $0$ & $\cdots$ & $0$ & $\cdots$ & $\cdots $ & $0$ & $\cdots$ & $0$ & $ /z^2$,\\
    \cline{1-8}
    \multicolumn{8}{c}{
      \hbox to 2.5\cwd{\upbracefill} 
      \hbox to 3.5\cwd{\upbracefill}
      \hbox to 2.5\cwd{\upbracefill}
      } \\
    \multicolumn{8}{c}{
      \hbox to 2.5\cwd{\hfill $g_{n_1}$ \hfill} 
      \hbox to 3.5\cwd{\hfill $g_{n_2}$ \hfill}
      \hbox to 2.5\cwd{\hfill $g_{n_3}$ \hfill}
      } \\
    \end{tabular}
\end{center}
with $n_1+n_2+n_3=n$. Some of the $n_i$'s in the product may be equal to each other.
The maximum order $n_k$ appearing in the partial sum, \emph{i.e.} the number of arguments of the multi-variate polynomial representing the sum, corresponds to the term where the $\ell$ initial (or final) $k_i$ indices in Eq. \eqref{eq:Gn} vanish. For instance, in the $\ell=2$ case examined above, this would be represented by the two diagrams:
\begin{center}
     \begin{tabular}{
     |>{\centering\arraybackslash}m{\cwd}
     |>{\centering\arraybackslash}m{\cwd}
     |>{\centering\arraybackslash}m{\cwd}
     |>{\centering\arraybackslash}m{\cwd}
     |>{\centering\arraybackslash}m{\cwd}
     |>{\centering\arraybackslash}m{\cwd}
     |>{\centering\arraybackslash}m{\cwd}
     |>{\centering\arraybackslash}m{\cwd}
     | c }
    \cline{1-8}
    $0$ & $0$ & $0$ & $\cdots$ & $\cdots$ & $\cdots $ & $\cdots$ & $0$ & $ /z^2$,\\
    \cline{1-8}
    \multicolumn{8}{c}{
      \hbox to 1.25\cwd{\upbracefill} 
      \hbox to 1.25\cwd{\upbracefill}
      \hbox to 6\cwd{\upbracefill}
      } \\
    \multicolumn{8}{c}{
      \hbox to 1.25\cwd{\hfill $g_{1}$ \hfill} 
      \hbox to 1.25\cwd{\hfill $g_{1}$ \hfill}
      \hbox to 6\cwd{\hfill $g_{n-2}$ \hfill}
      } \\
    \end{tabular}
  \end{center}
  and
  \begin{center}
    \begin{tabular}{
    |>{\centering\arraybackslash}m{\cwd}
    |>{\centering\arraybackslash}m{\cwd}
    |>{\centering\arraybackslash}m{\cwd}
    |>{\centering\arraybackslash}m{\cwd}
    |>{\centering\arraybackslash}m{\cwd}
    |>{\centering\arraybackslash}m{\cwd}
    |>{\centering\arraybackslash}m{\cwd}
    |>{\centering\arraybackslash}m{\cwd}
    | c }
   \cline{1-8}
   $0$ & $\cdots$ & $\cdots$ & $\cdots $ & $\cdots$ & $0$ & $0$ & $0$ & $ /z^2$,\\
   \cline{1-8}
   \multicolumn{8}{c}{
     \hbox to 6\cwd{\upbracefill} 
     \hbox to 1.25\cwd{\upbracefill}
     \hbox to 1.25\cwd{\upbracefill}
     } \\
   \multicolumn{8}{c}{
     ~\hbox to 6\cwd{\hfill $g_{n-2}$ \hfill} 
     \hbox to 1.25\cwd{\hfill ~~$g_{1}$ \hfill}
     \hbox to 1.25\cwd{\hfill ~~~$g_{1}$ \hfill}~~~
     } \\
   \end{tabular}
  \end{center}
both corresponding to the contribution $(g_1)^2 g_{n-2}$. In the general case, diagrams of this kind give rise to the contribution $(g_1)^\ell g_{n-\ell}$. The most general contribution to the $\ell$-th partial sum is thus a multi-variate monomial in the $g$'s of the form:
\begin{equation}
  C(\bm g_{n-l},\bm j_{n-l})=(g_1)^{j_1}(g_2)^{j_2} \cdots (g_{n-\ell})^{j_{n-\ell}},
\end{equation}
where $\bm g_{n-\ell}=\{g_1,g_2,\cdots g_{n-\ell}\}$ and $\bm j_{n-l}= \{j_1,j_2,\cdots j_{n-\ell}\}$ is an array of $n-\ell$ \emph{non-negative} integers satisfying the constraints:
\begin{equation}
  \begin{aligned}\label{eq:j-constraints}
    j_1+j_2+\cdots j_{n-\ell} &= \ell+1 \\
    j_1+2j_2+\cdots (n-\ell)j_{n-\ell} &=n,
  \end{aligned}
\end{equation}
and one or more of the $j_k$'s may vanish. The multiplicity $N(\bm j_{n-l})$ of the $C(\bm g_{n-l},\bm j_{n-l})$ monomial is equal to the number of ways a set of $\ell+1$ elements grouped in subsets of $\{j_1,j_2,\cdots j_{n-\ell}\}$ equal elements (some of the $j$'s may vanish), can be partitioned into $\ell+1$ boxes. Simple combinatorics gives:
\begin{equation}
   N(\bm j_{n-\ell}) = \frac{(\ell+1)!}{j_1! j_2! \cdots j_{n-\ell}!}.
\end{equation}
We conclude that Eq. \eqref{eq:Gn} can be put into the form:
\begin{multline}\label{eq:Gn-1}
    G_n(z) = \sum_{\ell=0}^{n-1}\frac{1}{z^\ell} \sum_{j_1j_2\cdots j_{n-\ell}} \frac{(\ell+1)!}{j_1! j_2! \cdots j_{n-\ell}!} \\ \times (g_1)^{j_1}(g_2)^{j_2} \cdots (g_{n-\ell})^{j_{n-\ell}},
\end{multline}
where the multiple sum is restricted to the $j$'s subject to the constraints in Eqs. \eqref{eq:j-constraints}. This multiple sum coincides with the definition of the \emph{ordinary Bell polynomial} \cite{Rebenda:2019} of order $(n,\ell+1)$, $\mathcal B_{n,\ell+1}(\bm g_{n-\ell})$ \cite{Bnote}. Eqs. (\ref{eq:lambda-bar}-\ref{eq:Gn}) can thus be cast into the form:
\begin{align}\label{eq:GnB}
  G_{n}(z) =\sum_{l=1}^{n}z^{-l+1}\mathcal{B}_{nl}\bigl (\mathbf{g}_{n-l+1}(z) \bigr ).
\end{align}
By extracting from the Laurent expansion of Eq. \eqref{eq:GnB} the terms of order one and zero and equating them to
$\lambda_{n}^{\circ}$, and $\dot{\lambda}_{n}^{\circ}$, respectively, as discussed before, one gets:
\begin{equation}\label{eq:lambda-0-Bell}
  \begin{aligned}
    \lambda_{n}^{\circ}
    & =\sum_{l=1}^{n}\frac{1}{l!}\mathcal{B}_{nl}^{(l)} \\
    {\dot\lambda}_{n}^{\circ}
    &= \sum_{l=1}^{n}\frac{1}{(l-1)!}\mathcal{B}_{nl}^{(l-1)},
  \end{aligned}
\end{equation}
where $\mathcal{B}_{nl}^{(k)} =\left.\frac{d^{k}}{dz^{k}} \mathcal{B}_{nl} \bigl(\mathbf{g}_{n-l+1}(z)\bigr)\right|_{z=0}$. These derivatives can be expressed as linear combinations of multiple derivatives of the $g_{n}$'s, $g_{n}^{(k)}= \left. \frac{d^{k}}{dz^{k}}g_{n}(z) \right|_{z=0}$,
using a multi-variate extension of the Fa\`a di Bruno formula \cite{Hardy:2006}, involving again Bell's polynomials. In practice, the coefficients of these linear combinations quickly become so complex that they can only be handled through symbolic manipulation systems, which would be more profitably used to obtain the result by direct differentiation. In any case, the derivatives of the $g_{n}$'s can be expressed in terms of the complete homogeneous symmetric polynomials \cite{Macdonald:1995} of the inverse excitation energies, $X_n=\mathcal{E}_{n}^{-1}$,
\begin{widetext}
\begin{equation}\label{eq:CompleteHomoPoly}
        h_{l}(X_{1},\cdots X_{n})
        =\sum_{1\le k_{1} \cdots k_n \le l}X_{k_{1}} \cdots X_{k_{n}}
        = \frac{1}{l!} \frac{d^{l}}{dz^{l}} \left(\frac{1}{1-zX_{1}}\cdots\frac{1}{1-zX_{n}}\right)_{z=0}. 
\end{equation}
We have therefore:
\begin{equation} \label{eq:g(l)n}
    \begin{aligned}
    g_{n}^{(l)} & =\left.\frac{d^{l}}{dz^{l}}g_{n}(z)\right|_{z=0} 
     =\frac{d^{l}}{dz^{l}}\left.\sideset{}{'}\sum_{k_{1}k_{2}\cdots k_{n-1}} \frac{\mathcal{W}_{0k_{n-1}} \mathcal{W}_{k_{n-1}k_{n-2}} \cdots \mathcal{W}_{k_10}}{(\mathcal{E}_{k_{n-1}}+z)
     \cdots(\mathcal{E}_{k_{1}}+z)} \right|_{z=0} \\& =l!(-)^{n-1} \sideset{}{'}\sum_{k_{1}k_{2}\cdots k_{n-1}}\frac{\mathcal{W}_{0k_{n-1}}\mathcal{W}_{k_{n-1}k_{n-2}}\cdots
     \mathcal{W}_{k_10}}{\mathcal{E}_{k_{n-1}}\cdots\mathcal{E}_{k_{1}}}h_{l}\left(\mathcal{E}_{k_{1}}^{-1},\cdots\mathcal{E}_{k_{n-1}}^{-1}\right).
    \end{aligned}
\end{equation}

The box below, Eqs. \eqref{eq:summary-1}, summarizes the formulas for the calculation of the various terms in the perturbative expansion of the GS energy of the Hamiltonian, Eq. \eqref{eq:H-split} to arbitary order: Eqs. \eqref{eq:kappa-perturbative}, \eqref{eq:gamma-naught-recursion}, and \eqref{eq:lambda-0-Bell}.
  \begin{equation} \label{eq:summary-1}
    \boxed{
    \begin{array}{rclcrclcrcl}
      E_{0} &=& \mathcal E_0 + \varepsilon_1  + \cdots \varepsilon_n+ \cdots, &\quad&
      \gamma^\circ_n &=& \lambda^\circ_n -\sum_{k=1}^{n-1} \frac{n-k}{n}\gamma^\circ_{n-k}\lambda^\circ_k, &\quad&
      \lambda^\circ_n & = & \sum_{l=1}^{n}\frac{1}{l!}\mathcal{B}_{nl}^{(l)}, \\
      \varepsilon_n &=& (-)^{n+1}{\dot\gamma}^\circ_{n}, &\quad& {\dot \gamma}^\circ_n &=& \dot{\lambda}_{n}^{\circ}-\sum_{k=1}^{n-1}\frac{n-k}{n}\left(\dot{\gamma}_{n-k}^{\circ}\lambda_{k}^{\circ}+\gamma_{n-k}^{\circ}\dot{\lambda}_{k}^{\circ}\right), &\quad&
      {\dot\lambda}^\circ_n &=& \sum_{l=1}^{n}\frac{1}{(l-1)!}\mathcal{B}_{nl}^{(l-1)}
    \end{array}
    }
  \end{equation}
\end{widetext}
These equations are easily implemented in any symbolic manipulation package. A simple Mathematica \cite{Mathematica} code, named \texttt{TuMiTurbi.nb}, is available as Supplemental Material / Ancillary File. The box below, Eqs. \eqref{eq:summary-2}, reports the first six terms in the perturbative expansion of the GS energy, as obtained from this code.
Note the difference between $g^k_l= (g_l \bigr )^k$ and $g^{(k)}_l=\frac{d^k g_l}{d z^k}$. These results are in agreement with those obtained in Ref. \onlinecite{Bracci2012} from a different method based on gauge invariance. \texttt{TuMiTurbi.nb} also provides the explicit expressions for the perturbative corrections in terms of the familiar sums over excited states, in a slightly awkward, but perfectly recognizable, form.

\begin{equation}\label{eq:summary-2}
    \boxed{
      \begin{aligned}
        \quad \varepsilon_1 &= g_1 \\
        \varepsilon_2 &= -g_2 \\
        \varepsilon_3 &= g_3+g_1 g_2' \\
        \varepsilon_4 &= -g_4-g_2 g_2'-g_1 g_3'-\frac{1}{2} g_1^2 g_2'' \\
        \varepsilon_5 &= g_5+g_3 g_2'+g_1 \left(g_2'\right){}^2+g_2 g_3'+g_1 g_4' \\ & \qquad +g_1 g_2 g_2''  +\frac{1}{2} g_1^2 g_3''+\frac{1}{6} g_1^3 g_2^{(3)} \\
        \varepsilon_6 &= -g_6 -g_4 g_2' -g_2 \left(g_2'\right)^2 -g_3 g_3' -2 g_1 g_2' g_3' \quad \\ &\qquad -g_2 g_4'
         -g_1 g_5'-\frac{1}{2} g_2^2 g_2''-g_1 g_3 g_2'' \\
        &\qquad -\frac{3}{2} g_1^2 g_2' g_2''  -g_1 g_2 g_3'' -  \frac{1}{2} g_1^2 g_4'' \\
        &\qquad -\frac{1}{2}
        g_1^2 g_2 g_2^{(3)}-\frac{1}{6} g_1^3 g_3^{(3)}-\frac{1}{24} g_1^4 g_2^{(4)}
      \end{aligned}
    }
\end{equation}


\section{The classical-quantum mapping} \label{sec:ClaQua}
In order to proceed further and establish a stochastic interpretation of the perturbative series, Eq. \eqref{eq:kappa-perturbative}, we consider a classical system of $N$ interacting particles, whose coordinates are denoted by $\bm R=\{ \bm r_1, \bm r_2, \cdots \bm r_N \}\in \mathbb{R}^{3N}$ and whose dynamics is described by a random walk satisfying the overdamped Langevin equation:
\begin{equation}
  \label{eq:Langevin}
  \begin{aligned}
    \bm R_{n+1} &= \bm R_n + \epsilon \bm{\mathcal{F}}(\bm R_n) +d\bm W_n, \\
    \bm{\mathcal{F}}&=-\frac{\partial\mathcal U(\bm R)}{\partial\bm R},
  \end{aligned}
\end{equation}
where  $\mathcal U(\bm R)$ is a many-body potential, $d\bm W_n$ is the differential of a Wiener process with variance $\bigl \langle \left ( d\bm W_n \right )^2 \bigr\rangle = 2\epsilon$, and the subscript $n$ is a discrete-time index corresponding to a discretization step $\epsilon$. In the continuous ($\epsilon\to 0$) limit, the probability density for the walker $\bm R$, $\mathsf P(\bm R,\tau)$, satisfies the Fokker-Planck (FP) equation \cite{Parisi:1981,Baroni:1999a}:
\begin{equation}
  \frac{\partial \mathsf P(\bm R,\tau)}{\partial \tau} =
  \frac{\partial^2 \mathsf P(\bm R,\tau)}{\partial \bm R^2} -
  \frac{\partial}{\partial\bm R} \cdot
  \bigl ( \bm{\mathcal{F}}(\bm R)\mathsf P(\bm R,\tau) \bigr ). \label{eq:Fokker-Planck}
\end{equation}
It is easily checked that ${\mathsf P}^\circ (\bm R) \propto \mathrm{e}^{-{\mathcal U}(\bm R)}$ is a stationary solution of the FP equation, Eq. \eqref{eq:Fokker-Planck}. We will shortly see that, under rather general conditions, this stationary solution is unique. To this end, let us introduce two auxiliary \emph{wavefunctions} defined as:
\begin{align}
  \Phi_0(\bm{R}) &= \sqrt{\mathsf{P}^\circ(\bm{R})} \propto  \mathrm{e}^{-{\mathcal U}(\bm R)/2},  \label{eq:Phi0}\\
  \Phi({\bm R},\tau) &= \mathsf{P}({\bm R},\tau)/ \Phi_0(\bm R). \label{eq:Phitau}
\end{align}
It is easy to verify that $\Phi({\bm R},\tau)$ satisfies the (imaginary-) time-dependent Schr\"odinger equation:
\begin{align}
  \frac{\partial\Phi(\bm{R},\tau)}{\partial\tau} &=-\widehat{\mathcal H} \Phi(\bm{R},\tau), \text{ where}\\
  \widehat{\mathcal H} &= -\frac{\partial^2 }{\partial \bm{R}^2} + \mathcal V(\bm{R}), \text{ and} \label{eq:AuxilaryHamiltonian} \\
  \mathcal V(\bm{R}) &= \frac{1}{4} \mathcal F(\bm{R})^2 - \frac{1}{2} \Delta \mathcal U(\bm{R}), \nonumber \\
&=\frac{\Phi''_0(\bm{R})}{\Phi_0(\bm{R})},
\label{eq:Veff}
\end{align}
where $\Phi''_0(\bm{R})=\frac{\partial^2}{\partial\bm R^2} \Phi_0(\bm R)$. Eqs. (\ref{eq:AuxilaryHamiltonian}-\ref{eq:Veff}) imply that $\Phi_0$, Eq. \eqref{eq:Phi0}, is an eigenfunction of the Hamiltonian, Eq. \eqref{eq:AuxilaryHamiltonian}, with zero eigenvalue. If $\mathcal U(\bm{R})$, Eq. \eqref{eq:Langevin}, is everywhere finite, then
$\mathsf{P}^\circ(\bm{R})$ and $\Phi_0(\bm{R})$ are nodeless, and the latter is the non-degenerate ground state of the Hamiltonian, Eq. \eqref{eq:AuxilaryHamiltonian} \cite{Feynman:1972}. As a consequence, all the excited states have strictly positive energies, and therefore $ \lim_{\tau\to\infty} \Phi(\bm{R},\tau) \propto \Phi_0(\bm{R})$ and
$ \lim_{\tau\to\infty} \mathsf{P}(\bm{R},\tau) = \mathsf{P}^\circ(\bm{R})$,
irrespective of the initial conditions, \emph{i.e.} $\mathsf{P}^\circ(\bm{R})$ is the unique equilibrium solution of the FP equation, Eq. \eqref{eq:Fokker-Planck}.

The FP equation, Eq. \eqref{eq:Fokker-Planck}, is first-order in time, reflecting the Markovian character of the Langevin process, Eq. \eqref{eq:Langevin}. This entails that its solution, $\mathsf P(\bm R,\tau)$, is uniquely determined by the corresponding initial condition, $\mathsf P(\bm R,0)$. Linearity in turn implies that $\mathsf P(\bm R,\tau)$ is the convolution of $\mathsf P(\bm R,0)$ with a Green's function, $\Pi(\bm R, \bm R'; \tau-\tau')$,
which is to be interpreted as the conditional probability density for the walker to be found at position $\bm R$ at time $\tau$, given that it was found at position $\bm R'$ at time $\tau'$:
\begin{equation}
  \mathsf P(\bm R,\tau) = \int \Pi(\bm R, \bm R'; \tau) \mathsf P(\bm R',0) d\bm R'.
\end{equation}
A similar relation holds for the propagation of the associated quantum wavefunction:
\begin{equation}
  \Phi(\bm R,\tau) = \int \mathcal G(\bm R, \bm R'; \tau) \Phi(\bm R',0) d\bm R', \label{eq:PhiProp}
\end{equation}
where $\mathcal G(\bm R, \bm R'; \tau) = \langle \bm R | \mathrm e^{-\widehat{\mathcal H}\tau} | \bm R' \rangle $ is the imaginary-time propagator of the auxiliary quantum system. By inserting Eq. \eqref{eq:Phitau} into Eq. \eqref{eq:PhiProp}, one gets:
\begin{align}
  \Pi(\bm R, \bm R'; \tau) = \Phi^\circ(\bm R) \mathcal G(\bm R, \bm R'; \tau) / \Phi^\circ(\bm R'). \\ \nonumber
\end{align}

If the system is initially at equilibrium, $\mathsf P(\bm R,0)=\mathsf P^\circ(\bm R)$, the time average of any function of the walker's coordinates, $\mathcal A(\bm{R})$,
\begin{align} \label{eq:Abar}
  \bar{\mathcal A}_{\mathcal T} = \frac{1}{\mathcal T} \int_0^{\mathcal T} \mathcal A \bigl ( \bm R(\tau) \bigr ) d\tau,
\end{align}
is a stochastic variable whose expectation is:
\begin{equation} \label{eq:<A>}
  \begin{aligned}
    \langle \bar{\mathcal A}_{\mathcal T} \rangle_{\scriptscriptstyle RW}
      &= \langle \mathcal A \rangle \\
        &\doteq \int  \mathcal A(\bm{R}) \mathsf{P}^\circ(\bm{R}) d\bm{R} \\
        & \equiv \langle \Phi_0 | \widehat{\mathcal A} | \Phi_0 \rangle,
  \end{aligned}
\end{equation}
and whose variance is:
\begin{equation} \label{eq:EH}
\begin{aligned}
  \mathsf{var} \bigl ( \bar{\mathcal A}_{\mathcal T} \bigr ) &= \frac{1}{\mathcal T^2}
  \left \langle
    \left (
      \int_0^{\mathcal{T}} \Delta \mathcal A(\tau) d\tau
    \right )^2
  \right \rangle_{\scriptstyle RW} \\
  &= \frac{2}{\mathcal T} \int_0^{\mathcal T} \bigl \langle \Delta \mathcal A(\tau) \Delta \mathcal A(0) \rangle_{\scriptscriptstyle RW} \left ( 1-\frac{\tau}{\mathcal T}\right ) d\tau \\
  &\sim \frac{2}{\mathcal T} \int_0^\infty \bigl \langle \Delta \mathcal A(\tau) \Delta \mathcal A(0) \rangle_{\scriptscriptstyle RW} d\tau,
\end{aligned}
\end{equation}
where
\begin{equation}
  \Delta \mathcal A(\tau) =  \mathcal{A} \bigl (\bm{R}(\tau) \bigr ) - \langle \mathcal{A} \rangle, \label{eq:DeltaA}
\end{equation}
$\langle\cdot\rangle_{\scriptscriptstyle RW}$ indicates an equilibrium average over the random walk,
and the last relation in Eq. \eqref{eq:EH} holds in the $\mathcal T\to\infty$ limit when $\int_0^\infty \langle \Delta \mathcal A(\tau) \Delta \mathcal A(0) \rangle_{\scriptscriptstyle RW} \tau d\tau < +\infty$. Notice the similarity between the expression for the variance for the time average of a function of the walker's coordinates, Eq. \eqref{eq:EH}, and the Einstein-Helfand expression for transport coefficients \cite{Einstein:1905,Helfand:1960,Baroni:2020,Grasselli:2021}. Eq. \eqref{eq:EH}, as well as the related equivalence between the Green-Kubo and Einstein-Helfand expressions for transport coefficients, is a direct consequence of the fact that the variance of the average of $N$ of stochastic variables (the integral in Eq. \ref{eq:Abar}) is equal to the sum of the all the elements of the covariance matrix divided by $N^2$, which for independent equally distributed variables results in the familiar law of large numbers.

\begin{widetext}
If the probability density for the walker's coordinates at $\tau_1=0$ is stationary, $\mathsf P(\bm R,0)=\mathsf P^\circ(\bm R)$, the joint probability density for the walker to be found at positions $\bm R_1, \bm R_2, \cdots \bm R_{n}$ at times $\tau_1, \tau_2,\cdots \tau_{n}$ is:
  \begin{multline}
    \mathsf P_n(\bm R_{n}, \tau_{n}; \bm R_{n-1} \tau_{n-1}; \cdots ;\bm R_1,\tau_1) = \Pi(\bm R_{n}, \bm R_{n-1}; \tau_{n}-\tau_{n-1}) \times \\ \Pi(\bm R_{n-1}, \bm R_{n-2}; \tau_{n-1}-\tau_{n-2}) \times \cdots \Pi(\bm R_{2}, \bm R_1; \tau_2-\tau_1) \mathsf P^\circ(\bm R_1).
  \end{multline}
The time correlation function of a function of the local coordinates, $\mathcal A(\bm R)$, reads therefore:
\begin{equation} \label{eq:Acorr}
  \begin{aligned}
    \langle \Delta\mathcal A(\tau) \Delta\mathcal A(0) \rangle_{\scriptscriptstyle RW} &= \int \mathsf P_2(\bm R_{2}, \tau; \bm R_1, 0 ) \Delta \mathcal A(\bm R_2) \Delta \mathcal A(\bm R_1) d\bm R_2 d\bm R_2 \\
    &= \int \mathsf \Phi_0(\bm R_{2}) \Phi_0(\bm R_1) \mathcal G(\bm R_2, \bm R_1; \tau) \Delta \mathcal A(\bm R_2) \Delta \mathcal A(\bm R_1) d\bm R_1 d\bm R_2 \\
    &= \sum_{n>0}|\mathcal A_{0n}|^2 \mathrm{e}^{-\mathcal E_n\tau},
  \end{aligned}
\end{equation}
\end{widetext}
where $\Phi_n$ and $\mathcal E_n$ indicate the eigenpair of the $n$-th excited state of the Hamiltonian, Eq. \eqref{eq:AuxilaryHamiltonian}, ${\mathcal A}_{0n} = \langle \Phi_0 | \widehat{ \mathcal A} | \Phi_n \rangle$, and the GS energy, $\mathcal E_0$, is assumed to vanish.
By combining Eq. \eqref{eq:EH} with Eq. \eqref{eq:Acorr}, we arrive at an expression for the variance of the time average of a function of the walker's coordinates in terms of a spectral sum for the associated quantum system:
\begin{align}
  \mathsf{var} \bigl ( \bar{\mathcal A}_{\mathcal T} \bigr ) \sim \frac{2}{\mathcal T} \sum_{n>0} \frac{|\mathcal A_{0n}|^2}{\mathcal E_n}.
\end{align}

\section{Stochastic perturbation theory} \label{sec:StochPert}
The approach to perturbation theory presented in Sec. \ref{sec:RalSchr} applies to any Hamiltonian that can be split as in Eq. \eqref{eq:H-split}. When both the complete and unperturbed Hamiltonians of an $N$-body system are sums of a kinetic and a local, possibly non-separable, potential term, the GS wavefunctions are nodeless \cite{Feynman:1972,nodeless} and the unperturbed quantum problem can be mapped onto a classical diffusion one, such that the perturbative expansion can be given a nice and insightful stochastic interpretation.

Let us denote by $\bm R=\{ \bm r_1, \bm r_2, \cdots \bm r_N \}\in \mathbb R^{3N}$ the coordinates of the system and by
\begin{equation}
  \begin{aligned}
    \widehat H &= -\frac{1}{2}\frac{\partial^2}{\partial \bm R^2} + V(\bm R) \\
    \widehat{\mathcal H} &=-\frac{1}{2}\frac{\partial^2}{\partial \bm R^2} + \mathcal V(\bm R), \\
    \mathcal W(\bm R) &= V(\bm R) - \mathcal V(\bm R)
  \end{aligned}
\end{equation}
the complete and unperturbed Hamiltonians, respectively. The eigenvalue equation, Eq. \eqref{eq:eigenvalue}, gives:
\begin{equation}
  \mathcal V(\bm R) = \mathcal E_0 + \frac{1}{2}\frac{\Phi''_0(\bm R)}{\Phi_0(\bm R)} \smallskip
\end{equation}
where $\Phi_0$ is the unperturbed GS wavefunction.
If one assumes
$\mathcal E_0=0$, then
\begin{equation} \label{eq:W}
  \begin{aligned}
    \mathcal W(\bm R) &= -\frac{1}{2}\frac{\Phi''_0(\bm R)}{\Phi_0(\bm R)} + V(\bm R) \\
    &= \left ( \widehat{H} \Phi_0(\bm R) \right )  \Big / \Phi_0(\bm R).
  \end{aligned}
\end{equation}
In the quantum Monte Carlo parlance, the perturbing potential, $ \mathcal W(\bm R) $, Eq. \eqref{eq:W}, is usually dubbed the \emph{local energy}.

\begin{widetext}
A stochastic interpretation of the perturbative expansion is obtained by replacing the multiple sum over intermediate Hamiltonian eigenstates leading from Eq. \eqref{eq:lambda-def} to \eqref{eq:lambda-Sum_k} with a multiple integral over intermediate positions, reading:
  \begin{equation} \label{eq:lambda-stochastic}
    \begin{aligned}
      \lambda_{n}(\tau) & = \int_{0}^{\tau}d\tau_{n}\int_{0}^{\tau_{n}}d\tau_{n-1}\cdots\int_{0}^{\tau_{2}}d\tau_{1}
      \int d\bm{R}_n d\bm{R}_{n-1} \cdots d\bm{R}_1 \Phi_{0}(\bm{R}_n)\mathcal{W}(\bm{R}_{n})
      \mathcal{W}(\bm{R}_{n-1})\cdots \mathcal{W} (\bm{R}_{1}) \\
      & \hspace{69pt} \times \mathcal{G}(\bm{R}_{n},\bm{R}_{n-1};\tau_n-\tau_{n-1}) \mathcal{G}(\bm{R}_{n-1},\bm{R}_{n-2};\tau_{n-1}-\tau_{n-2}) \cdots \mathcal{G}(\bm{R}_{2},\bm{R}_{1};\tau_2-\tau_{1}) \Phi_{0}(\bm{R}_1) \\
      &= \int_{0}^{\tau}d\tau_{n}\int_{0}^{\tau_{n}}d\tau_{n-1}\cdots\int_{0}^{\tau_{2}}d\tau_{1} \int d\bm{R}_n d\bm{R}_{n-1} \cdots d\bm{R}_1 \mathcal{W}(\bm{R}_{n})
      \mathcal{W}(\bm{R}_{n-1})\cdots \mathcal{W} (\bm{R}_{1}) \\
      & \hspace{281pt}\times \mathsf P_n(\bm R_{n}, \tau_{n}; \bm R_{n-1} \tau_{n-1}; \cdots ;\bm R_1,\tau_1) \\
      &= \int_{0}^{\tau}d\tau_{n}\int_{0}^{\tau_{n}}d\tau_{n-1}\cdots\int_{0}^{\tau_{2}}d\tau_{1} \langle \mathcal{W}(\tau_{n}) \mathcal{W}(\tau_{n-1})\cdots \mathcal{W} (\tau_{1})\rangle_{\scriptscriptstyle RW} \\
      &= \frac{1}{n!}
      \bigl \langle \mathcal{S}(\tau)^n
      \bigr\rangle_{\scriptstyle RW},
    \end{aligned}
  \end{equation}
where $\mathcal S(\tau)=\int_0^\tau \mathcal W(\tau')d\tau'$
can be thought of as an effective action \cite{CarleoPhD}. The $\mu$'s, $\mu_n(\tau)=n!\lambda_n(\tau)$, Eq. \eqref{eq:lambda-def}, are thus the (raw) moments of the effective action, and the various perturbative corrections in Eq. \eqref{eq:kappa-perturbative} are derivatives of the corresponding cumulants. When $\tau$ is larger than the local-energy ($\mathcal W$) autocorrelation time, $\tau_{\scriptscriptstyle \mathcal{W}}$,
$\mathcal S(\tau)$ is the sum of $\mathcal N \approx \tau/\tau_{\scriptscriptstyle \mathcal{W}}$
quasi-independent stochastic variables, so that its cumulants are proportional to $\mathcal N$, and therefore to $\tau$, making the large-time limit of their derivatives well defined.
\end{widetext}
\section{Reptation Quantum Monte Carlo} \label{sec:RQMC}
The most basic of all the stochastic approaches to quantum mechanics is likely \emph{variational quantum Monte Carlo} (VMC), whereby one aims to estimate the GS energy of a system as the expectation value of the Hamiltonian with respect to a suitably identified approximate wave-function, $\Phi_0(\bm R)$:
\begin{equation}
  \begin{aligned}
    E_0 &\approx \langle \Phi_0|\widehat H | \Phi_0 \rangle \\
    &\doteq \int \mathcal W(\bm R) \Phi_0(\bm R)^2 d\bm R,
  \end{aligned}
\end{equation}
where $\mathcal W(\bm R)$ is given by Eq. \eqref{eq:W}. This is conveniently achieved by sampling $\mathcal W(\bm R)$ along a random walk generated by the Langevin equation, Eq. \eqref{eq:Langevin}, with $\mathcal U(\bm R)=-2\log\Phi_0(\bm R)$, using Eqs. (\ref{eq:<A>}-\ref{eq:EH}) with $\mathcal A=\mathcal W$.

The classical-quantum mapping presented in Sec. \ref{sec:ClaQua} permits to interpret $\Phi_0$ as the GS wavefunction of the auxiliary Hamiltonian, $\widehat{\mathcal H}$, associated with the FP equation for the Langevin random walk. Of course, if $\Phi_0$ coincided with the exact wavefunction of our quantum system, $\widehat{\mathcal H}$ would coincide with the exact Hamiltonian, $\widehat H$. If this is not the case, it would be reasonable to treat the difference $\widehat H - \widehat{\mathcal H} = \widehat{\mathcal W}$ by perturbation theory. According to Eqs. (\ref{eq:gamma-recursion}-\ref{eq:kappa-perturbative}) and \eqref{eq:lambda-stochastic}, the first few corrections to the unperturbed ($\mathcal E_0=0$) energy read:
\begin{align}
  \varepsilon_1 &= \langle\mathcal W\rangle_{\scriptscriptstyle RW} \label{eq:e1VMC} \\
  \varepsilon_2 &= - \int_0^\infty \bigl \langle \Delta \mathcal W(\tau) \Delta \mathcal W(0) \rangle_{\scriptscriptstyle RW} d\tau \label{eq:taudef} \\
  &\doteq -\bigl\langle (\Delta \mathcal W)^2\bigr\rangle_{\scriptscriptstyle RW} \tau_{\scriptscriptstyle \mathcal W}, \label{eq:e2VMC}
\end{align}
where $\Delta\mathcal W$ is defined in analogy with Eq. \eqref{eq:DeltaA} and the local-energy auto-correlation time, $\tau_{\scriptscriptstyle\mathcal W}$, is actually defined by Eqs. (\ref{eq:taudef}-\ref{eq:e2VMC}). The first-order correction, Eq. \eqref{eq:e1VMC}, coincides with the VMC estimate of the GS energy. Eq. \eqref{eq:e2VMC} states that the information contained in the local-energy time series generated in a regular VMC simulation is sufficient to evaluate the second- (and, actually, higher\nobreakdash-) order correction(s) to the VMC estimate.

The stochastic interpretation of the higher-order terms, Eqs. \eqref{eq:Zdef} and \eqref{eq:lambda-stochastic}, allows one to formally sum the perturbative series up to infinite order as the expected value over the random walk of the exponential of the negative of the action:
\begin{align}
  \mathcal Z(\tau) &= \sum_{n=0}^\infty \frac{(-)^n}{n!} \bigl \langle \mathcal{S}(\tau)^n \bigr\rangle_{\scriptstyle RW} \\
  &= \left \langle \mathrm{e}^{-\mathcal{S}(\tau)} \right\rangle_{\scriptstyle RW}.
\end{align}
The expression given by Eq. \eqref{eq:E-dlog} for the GS energy reads therefore:
\begin{equation} \label{eq:RQMCenergy}
  E_0\sim \frac{\bigl\langle \mathcal W(\tau) \mathrm{e}^{-\mathcal S(\tau)}\bigr\rangle_{\scriptscriptstyle RW}}{\bigl\langle \mathrm{e}^{-\mathcal S(\tau)}\bigr\rangle_{\scriptscriptstyle RW}}.
\end{equation}

Neglecting action fluctuations, Eq. \eqref{eq:RQMCenergy} reduces to the usual VMC expression for the energy. These fluctuations could be accounted for by weighting the local energy with $\mathrm{e}^{-\mathcal S(\tau)}$, resulting in the \emph{pure-diffusion quantum Monte Carlo} scheme of Ref. \onlinecite{Caffarel1988}. The exponential dependence of the weights on the action and the extensive character of the latter, however, make this scheme unfit but for systems of very small size and not very efficient otherwise. Similar approaches, all derived from a Feynman-Kac expression for $\mathcal Z$ function in Eq. \eqref{eq:E-dlog}, are the \emph{variational path integral} method of Ref. \onlinecite{Ceperley1995}, later rebranded as \emph{path-integral ground state} \cite{Sarsa2000}, and RQMC \cite{Baroni:1999a,Baroni:1999b}. In all these methods, the effects of the weights are accounted for by sampling the space of random walks of length $\tau$,
$\bm X(\tau) = \{\bm R(\epsilon), \bm R(2\epsilon), \cdots \bm R(\tau=n\epsilon)\}$ according to a Metropolis algorithm \cite{Metropolis1953}. The distintive feature of RQMC is the way Monte Carlo moves are generated by letting the random walk (the \emph{reptile}) creep back and forth for a certain time according to the Langevin equation, \eqref{eq:Langevin}, and accepted or rejected according to a Metropolis test on the variation of the effective action determined by the move. Beside the energy, RQMC allows for an unbiased estimate of general local observables, as well as of their static and dynamic (in imaginary time) response functions. The algorithm is explained in full detail elsewhere  \cite{Baroni:1999a,Baroni:1999b}, and I feel that this a good place to stop.

\section{Conclusions} \label{sec:Conclusions}
The work presented in this paper is made of two independent parts, whose main link is their relation to the development of reptation quantum Monte Carlo in the late nineties. Indeed, this development was motivated by the observation that the leading correction to the variational estimate of a ground-state energy is determined by the Kubo-like formula given by Eq. \eqref{eq:taudef} and by the difficulty to generalize it to higher orders in any useful manner. Sometimes, insurmountable difficulties are fortunate, for RQMC has proven to be much more powerful than any approximate perturbative schemes ever could: besides the intrisically approximate character of perturbation theory, the main numerical limitation to a stochastic approach to it is the increasing numerical noise affecting the estimate of the action moments for increasing order and the ill-conditioned nature of the expression of cumulants in terms of moments, Eqs. (\ref{eq:kappa-recursion}-\ref{eq:gamma-recursion}), due to sign alternation. The first part of this work, Sec. \ref{sec:RalSchr} is to a large extent unrelated from the second, but for the fact that I have long been wondering how Eq. \eqref{eq:E-dlog}, which is the starting point of RQMC and of many other quantum stochastic simulation methods, could be used to streamline the derivaton of Raleigh-Schr\"odinger perturbation theory. I hope the present paper provides a nice, though not necessarily impactful, answer to this question.

\acknowledgements
I wish to thank Saverio Moroni for inspiring this work and for sharing with me the joys and pains of the development and early applications of RQMC. I am grateful to Luigi E. Picasso for teaching me the rudiments of quantum mechanics and of clean thinking, too long ago to remember. This paper would never have seen the light of day if Giovanni B. Bachelet had not insisted that I present its content at the CECAM workshop on \emph{Recent developments in quantum Monte Carlo}, held in Rome in October 2021 to honor Saverio's sixty-first birthday (the celebration of his sixieth birthday was cancelled because of the restrictions due to the outburst of the COVID-19 pandemic). Finally, I am grateful to Federico Grasselli, Paolo Pegolo, and Cyrus Umrigar for a critical reading of the manuscript and to PP for assisting my rather poor Mathematica coding. This work was partially supported by the European Commission through the \textsc{MaX} Centre of Excellence for supercomputing applications (grant number 824143) and by the Italian MUR, through the PRIN project \emph{FERMAT} (grant number 2017KFY7XF).

\bibliography{saverio}

\end{document}